\documentclass[reprint,aip,apl,twocolumn,color,psfig]{revtex4-1}
\usepackage{amsmath}
\usepackage{color}
\usepackage{amsfonts}
\usepackage{graphicx}

\begin{document}

\title{Switching and defect dynamics in multistable liquid crystal devices}



\author{A. Tiribocchi} 
\affiliation{Dipartimento di Fisica {\rm and} Sez.~INFN di Bari, Universit\`a di Bari, 70126 Bari Italy}

\author{G. Gonnella} 
\affiliation{Dipartimento di Fisica {\rm and} Sez.~INFN di Bari, Universit\`a di Bari, 70126 Bari Italy}

\author{D. Marenduzzo}
\affiliation{SUPA, School of Physics, University of Edinburgh, 
Edinburgh EH9 3JZ  UK}

\author{E. Orlandini}
\affiliation{Dipartimento di  Fisica {\rm and} Sez.~INFN di Padova, Universit\`a di Padova, 35131 Padova Italy}

\begin{abstract}
We investigate the switching dynamics of multistable nematic liquid crystal devices. In particular we identify a remarkably simple 2-dimensional (2D) device which exploits hybrid alignment at the surfaces to yield a bistable response. We also consider a 3-dimensional (3D) tristable nematic device with patterned anchoring, recently implemented in practice, and discuss how the director and disclination patterns change during switching.
\end{abstract}

\maketitle

Liquid crystals (LCs) are interesting soft materials with widespread technological applications, for instance liquid crystal displays (LCDs).
The archetype LCD device is the ``twisted nematic display'', commonly employed in flat panel monitors~\cite{chandrasekhar}. This device is easy to build in practice, but its functioning is not optimal. Firstly, it requires the constant application of an electric field for the whole time the device is to be in the ``on'' state. Secondly, standard twisted nematic LCD devices suffer from the so-called ``viewing angle problem''~\cite{view1dtn}, which refers to the limited range of angles from which the display is readable.
Both these shortcomings can be avoided at least conceptually. On one hand, the viewing angle range can be dramatically enhanced by building a ``multi-domain'' device, with the LC anchoring patterned on the device surface: this leads to the simultaneous presence of regions of left and right-handed twist, whose viewing angle characteristics differ and complement each other. Modelling plays an important role in optimising the delicate balance between the gain in viewing angle range and the loss in resolution due to the creation of defects between regions with different director orientation~\cite{kent,4dva,mdlcd1,Marenduzzo05}. On the other hand, bistable or multistable devices can retain memory of two or more distinct states, even after the electric field is switched off. These solved the problem of having to constantly use a large field to keep the device in the ``on'' state. Multistable devices are also interesting theoretically as they require the presence of a series of metastable states, with comparable free energy yet different defect topology and optical properties. How to realise this in practice is usually non-trivial. Two well known multistable devices proposed in the physical literature are the zenithal-bistable device~\cite{Denniston01}, and the tri-stable device experimentally demonstrated in~\cite{Kim02}. In both cases, defect dynamics is likely to be important~\cite{Denniston01,Kim02}.

Here we describe the physics behind the switching dynamics of two multistable and multidomain nematic devices. Our main result is that it is possible to design a remarkably simple 2-domain hybridly aligned nematic (HAN) cell (Fig.~1a). For a wide range of parameters we find two metastable states with competing free energy: one in which the LC pattern is made up by stripes with alternating splay-bend direction with defects in between, and another one in which the device becomes effectively a single domain with either clockwise or anticlockwise deformation. The dynamical schedule with which the electric field is switched on and off is crucial to ensure bistability. We also investigate a 3D tristable device similar to the one built in~\cite{Kim02} (Fig.~1b), and our results suggest that the disclination dynamics involved in the switching is strikingly non-trivial. \if{First, we show that the texture forced by the boundaries used in ~\cite{Kim02} is asymmetrical, and straight disclinations of topological charge 1/2 and 1 coexist in steady state. When a field is applied, the latter immediately unzip into a loop with half topological charge, to restructure later on when the field is large enough, and absorb onto the surface.}\fi Finally, we quantify the effect of backflow in the switching of the first of these devices and we show that it can change the range of voltages for which multistability occurs.

The physics of the LC device may be described by a Landau-de Gennes free energy dependent on a tensor order parameter, $Q_{\alpha\beta}$, whose largest eigenvalue (denoted by 2/3$q$) gives the magnitude of the local order. The free energy density $f$ is a sum of three terms. The first is a bulk contribution,
\begin{eqnarray}
\nonumber
\frac{A_0}{2}(1 - \frac {\gamma} {3}) Q_{\alpha \beta}^2 -
          \frac {A_0 \gamma}{3} Q_{\alpha \beta}Q_{\beta
          \gamma}Q_{\gamma \alpha}
+ \frac {A_0 \gamma}{4} (Q_{\alpha \beta}^2)^2,
\label{eqBulkFree}
\end{eqnarray}
where $A_0$ is a constant and $\gamma$ controls the magnitude of the ordering (Greek indices denote Cartesian components and summation over repeated indices is implied). A second term ${K}/{2} \left(\partial_{\gamma} Q_{\alpha \beta}\right)^2$ describes the free energy cost of distortions~\cite{chandrasekhar} where $K$ is an elastic constant. The last term, $-\frac{\varepsilon_a}{12\pi}E_{\alpha} Q_{\alpha \beta} E_{\beta}$, describes the interaction with the external electric field ${\bf E}$ where $\varepsilon_a >0$ is the dielectric anisotropy of the material.
The equation of motion for {\bf Q} is  $D_t{\bf Q}= \Gamma {\bf H}$, where $\Gamma$ is a collective rotational diffusion constant and $D_t$ is the material derivative for rod-like molecules~\cite{beris}. The molecular field $\mathbf{H}$ ensures that $\mathbf{Q}$ evolves towards a minimum of the free energy~\cite{beris}. The velocity field obeys a Navier-Stokes equation with a stress tensor generalised to describe LC hydrodynamics~\cite{beris,colin}. To solve the equations of motion, we used a hybrid lattice Boltzmann algorithm, see e.g.~\cite{hybrid}.
In what follows we express parameters in simulation units. To convert these to real units, one may assume to model a $\sim\mu$m thick device and a flow-aligning~\cite{chandrasekhar,beris} LC with rotational viscosity equal to 1 poise -- a typical value for e.g. 5CB. In this way, one force, time, and space unit may be mapped to 62.5 pN, 0.7 $\mu$s, and 0.025 $\mu$m respectively. The electric field may be quantified via the dimensionless number $e^2=27 \epsilon_a E_{\alpha}^2/(32\pi A_0\gamma)$. The simulation domains along the three coordinate axes are named $L_x$, $L_y$ and $L_z$ respectively (there are periodic boundaries along $\hat{x}$ and $\hat{y}$).

We begin by considering the two-domain device shown in Fig.~1a, with mixed anchoring, homeotropic -- along $z$ -- at the top ($z=L$) and homogeneous -- along y -- at the bottom ($z=0$) of the cell. A small pretilt is patterned to create a defect in the middle of the cell, where a region of clockwise splay-bend director rotation meets a mirror image with anticlockwise deformation. We call this zero-field free energy minimum the ``two-domain'' state. The geometry of this device is 2D, as there is translational invariance along $\hat{x}$. This device is a simple variant of the commonly used HAN cell, and it may be built by either chemical or topographical patterning. Our device contains coexisting homeotropic and homogeneous domains, which might lead to a switching in either direction, through the growth of one or the other domain: hence bistability is possible.
\begin{figure}
\centerline
{\includegraphics[width=8cm]{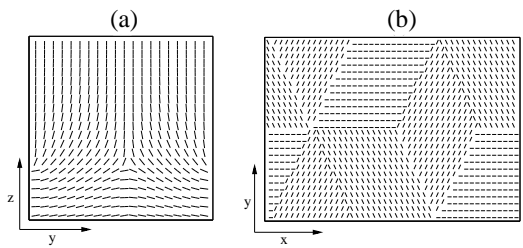}}
\caption{Geometry of the multidomain devices in this work. (a) 2-domain cell, with homeotropic  anchoring at the top plate and homogeneous anchoring with pretilt patterning at the bottom plate.
(b) Surface patterning used for the top and bottom boundary conditions
of the 3-domain tristable device.
}
\end{figure}
Fig.~2a-c shows the evolution of the director field in response to an electric field 
which is switched on and off first along $\hat{y}$, then along $\hat{z}$. 
After application of the field along $\hat{y}$, the disclinations separating 
neighbouring clockwise and anticlockwise domains annihilate and transform the two-domain device 
into a single domain one. This defect-free state is stable after the removal of the field, 
as the pretilt is not enough to bend the director field in the bulk. On the other hand, when a 
sufficiently strong field is applied along $\hat{z}$, defects reappear close to the surface, between regions with different pretilt. After switching off, 
elasticity alone is not enough to melt these defects away -- rather they migrate back to the 
centre of the sample and reconstitute the initial two-domain structure.
\begin{figure}
\centerline {\includegraphics[width=8cm]{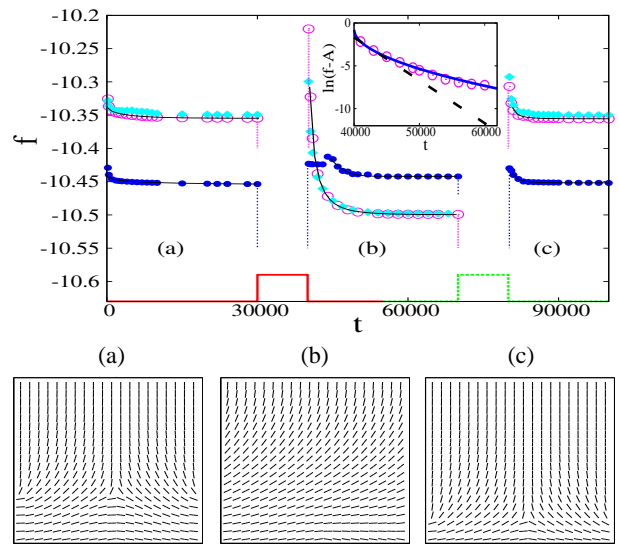}}
\caption{Free energy evolution (top panel) of the 2D device. 
Parameters were $A_0=1$, $\Gamma=0.33775$, $\gamma=3$, $L_y=L_z=40$, 
$e^2 \sim 0.06$, $K=0.08$ (open circles) and $K=0.04$ (filled circles).
All quantities are given in simulation units. 
Filled squares correspond to a simulation with backflow. The director field configuration of the corresponding stable states are reported in the bottom panel. 
The device is only bistable for $K=0.08$. The step function in the top panel is non-zero when the field is applied.
From $t=0$ up to $t=30000$ the system relaxes from the configuration depicted in Fig.~1 to a stable state, ((a) in the bottom panel). 
An electric field along $\hat{y}$ is switched on at $t=30000$ and off at $t=40000$; the system relaxes to a stable state, ((b) in the bottom panel,
snapshot at $t=70000$) characterized by a new free energy minimum.
At $t=70000$ an electric field along $\hat{z}$ is switched on, and off at $t=80000$. The relaxation process brings the system back to state (a) (with the defect in a slightly different position). The relaxation curves for the two cases with no backflow (open and filled circles) were fitted via a stretched exponential, $A+Be^{-\left(t/\tau\right)^{\beta}}$, (solid thick lines, see inset for relaxation to state (b) with $K=0.08$, where the dashed line is instead a fit to a simple exponential). 
For $K=0.08$, $\tau$ and $\beta$ are $\sim$ $3200$ and $0.72$ (a), $540$ and $0.52$ (b); $80$ and $0.39$ (c). For $K=0.04$, $\tau$ and $\beta$ are $\sim$ $5140$ and $0.51$ (a), $3460$ and $1.50$ (b); $360$ and $0.56$ (c). }
\end{figure}
The time evolution of the free energy during this electric field cycle is shown in Fig.~2 
(top panel, top set of curves) -- this shows that both the single domain and the two-domain states are metastable and close in free energy,
the main requirement for a bistable device. 

Changing the device size and the elastic constant affects both the switching on and off times and the bistability. 
E.g. relaxation times  after switching off of the horizontal  electric field, (see caption Fig.~2), scale like $L_z^2/K\Gamma$ (data not shown), as for common nematic devices~\cite{chandrasekhar}. 
The device is bistable only for small
size (e.g. $L_z<60$ for $K=0.08$ and other parameters as in the caption of Fig.~2) -- for 
larger devices the application of a field perpendicular to the boundary fails to turn the 
single domain state back into the two-domain one. Furthemore, bistability occurs for 
intermediate values of $K$ -- between about $0.06$ and $0.09$ for $L_z=50$, and other parameters 
as in Fig.~2 (Fig.~2, curves for $K=0.04$ with only one stable state). This is because when $K$ is too small, there is 
not enough elastic driving force to lead to defect annihilation during the first half of the 
voltage cycle, whereas when $K$ is too large the single domain state becomes too low in free 
energy with respect to the two-domain one, and the dynamics always gets 
stuck there. 

Backflow quantitatively affects the extent of the bistable 
regime. When it is neglected, the electric field range for which bistability 
is observed is overestimated. E.g., with no backflow, $K=0.08$ and other parameters as 
in Fig.~2, bistability requires $e^2>0.03$, whereas with backflow $e^2>0.05$ 
(filled squares in Fig.~2, top). 
This is because hydrodynamics decreases the chance of 
getting stuck into the two competing metastable states with different free energies.

The idea of using domains of different director orientations as seeds for multistable field-induced steady states has been used for the tristable device in~\cite{Kim02}. The relevant geometry is shown in Fig.~1b -- it consists of a hexagonal tiling of one or both boundary planes. In order to tile the whole boundary, three director orientations are needed. Via an electric field in the $xy$ plane, the authors of~\cite{Kim02} managed to stabilise each of the orientations used for the surface patterning. But what exactly is the switching dynamics of this multistable device? What is the LC texture corresponding to the zero-field case? And what is the role of defect motion for the device switching? Such questions cannot be easily answered experimentally -- yet are important to better control the device performance. Our numerical approach is ideal to fill in this gap and follow the director dynamics during switching (Fig.~3).
\begin{figure}
\centerline {\includegraphics[width=8.9cm]{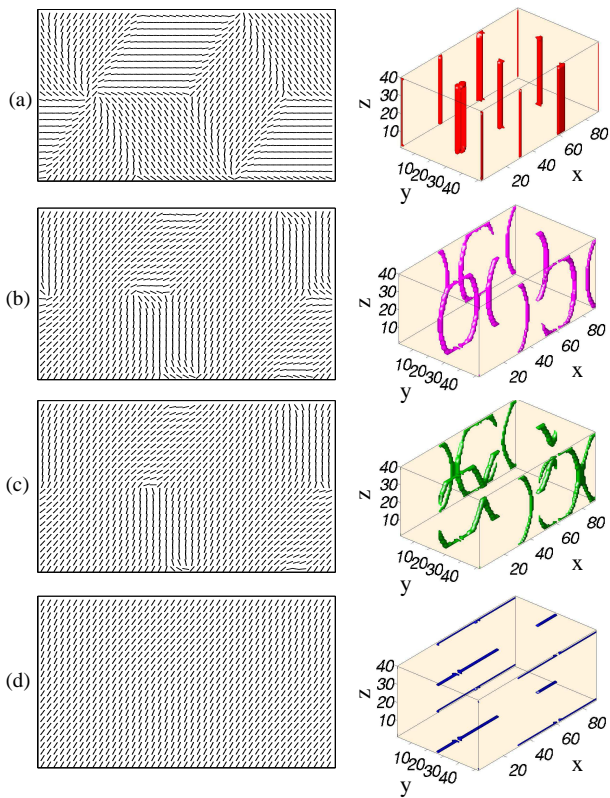}}
\caption{Evolution of the director field in the midplane (left column) and 
disclinations (right column) in the 3D tristable
device. Snapshots corresponds to time equal to (a) 100, (b) 3700, (c) 4600 and
(d) 6600 simulation units. Parameters are:
$L_x=90$, $L_y=52$, $L_z=40$, $K=0.08$, $e^2 = 0.0018$, and other parameters as in Fig.~2. The field is applied at $t=0$ and at an angle of $60^{\circ}$ with respect to $\hat{x}$.}
\end{figure}
For $e^2=0$, the surface patterning forces the formation of columnar straight disclinations -- of both integer and half-integer topological charge (Fig.~3a). This 3-domain state is metastable. When an electric field is applied, along one of the anchoring directions, the line of asters, which marks the point at which six different domains meet, unzips into a loop of half-integer topological charge. This dynamics is linked to a complex reshaping of the director field in the bulk (see Figs.~3b and 3c). Later on, the disclinations loop merge and reconstruct into an array of disclinations, which are now parallel to the boundary planes and pinned to the surface (Fig.~3d). This electric-field induced state is stabilised by the bulk elastic term even after switching off. 
Two important control parameters determining the physics of our tristable device are the ratio between domain and sample size, $d/L$, and the ``interface number'' $d/\sqrt{A_0/K}$ proportional to the ratio between the domain and defect core size. Briefly, additional simulations show that, if the sample is too thin, boundaries always restore the same state after switching off the inplane field, and multistability is lost. On the other hand, increasing the ``interface number'' would stabilise a disclination network in the bulk of the device -- however this would only become relevant for $d$ in the sub-$\mu$m range, difficult to reach in practice.

Thus, we studied the switching dynamics of multistable nematic LCs. These are important for applications, as they enable a device to function without the constant application of an electric field. We have shown that a 2D HAN cell, with homeotropic anchoring at one boundary and planar patterned surface anchoring at the other boundary provides a strikingly simple example of a bistable device. The simplicity of the geometry considered and of the electric field cycle should render this device easy to be built.
We also elucidated the role of defect dynamics in the switching of a tristable nematic device akin to that of~\cite{Kim02}, which was again stabilised by surface patterning. The spectacular defect restructuring in Fig.~3 should leave an observable signature in the optical patterns which can be recorded experimentally.

\end{document}